\documentstyle[aps,epsf]{revtex}  % DO NOT DELETE THIS LINE 

\begin{document}        %  DO NOT DELETE OR CHANGE THIS LINE

\baselineskip 14pt
\title{Higgs Compositeness from Top Dynamics and Extra Dimensions} 
\author{Bogdan A. Dobrescu} 
%  \vspace*{2mm}
\address{Fermi National Accelerator Laboratory \\
P.O. Box 500, Batavia, IL 60510, USA \thanks{e-mail
  address: bdob@fnal.gov}}
\maketitle              % Creates the title area, Do Not Remove

\vspace*{-3.3cm}
\noindent
{\small
\makebox[13.8cm][l]{hep-ph/9903407}
 FERMILAB-Conf-99/051-T \\ 
}
 \vspace*{2.8cm}

\centerline{March 18, 1999} 

\vspace*{.2cm}

\begin{abstract}
The fundamental Higgs doublet may be replaced in the Standard Model
by certain non-perturbative four-quark interactions,
whose effect is to induce a composite Higgs sector responsible 
for electroweak symmetry breaking. A simple composite two-Higgs-doublet 
model is presented. 
The four-quark interactions arise naturally if there are either extra spatial 
dimensions or larger gauge symmetries at a multi-TeV scale.
Some theoretical and phenomenological implications 
of these scenarios are discussed.
\end{abstract}   	

%%%%%%%%%%%%%%%%%%%%%%%%%%%%%%%%%%%%%%%%%%%%%%%%%%%%%%%%%%%%%%%
\section{Motivation}               % Introduction goes below.

The fundamental interactions observed so far in experiments
are the $SU(3)_C \times SU(2)_W \times U(1)_Y$ gauge interactions
acting on three generations of chiral fermions, measured at length scales
larger than $10^{-16}$ cm, and the universal gravitational 
interactions, measured at length scales larger than about 1 mm.
In addition, it is necessary that some new interactions 
break spontaneously the electroweak symmetry. 
In the Standard Model these new interactions involve a fundamental
Higgs doublet, which is a convenient but rather ad-hoc assumption.
It is troublesome that the Higgs mass is not controlled 
by gauge invariance.
The same remains true for the Supersymmetric Standard 
Model, where the $\mu$ term is not determined by the supersymmetry 
breaking scale. 
Although this problem may be cured by the presence of a new gauge group
whose breaking is linked to the transmission of supersymmetry 
breaking \cite{mu}, it remains desirable to investigate other origins of the
electroweak asymmetry of the vacuum.

It is remarkable that the electroweak symmetry may be spontaneously
broken without the need for new fields. This is possible because
the electroweak symmetry is embedded in the chiral symmetry 
of the quarks and leptons. Given that strongly coupled
four-fermion interactions induce chiral symmetry breaking \cite{njl},
it is possible to replace the fundamental Higgs doublet by 
certain four-quark operators \cite{condens,bhl}.
However, a computation of the $W$ and $Z$ masses in the large  
$N_c$ limit indicates that the electroweak breaking quark mass should
be of order 0.5 TeV in the absence of excessive fine-tuning.
It would be tempting then to consider a fourth generation of fermions, but
 this is disfavored by the current electroweak data 
(at the 99.8\% confidence level \cite{lang} in the case of degenerate 
fermions).

There is a simple solution to this puzzle: to introduce a seventh quark,
$\chi$, whose left and right components transform as the right-handed top 
quark, $t_R$, under the 
Standard Model gauge group. In this case, there is mass mixing between 
$\chi$ and $t$. The electroweak breaking $\bar{\chi}_R t_L$ mass induced 
by four-quark interactions can be of order 0.5 TeV, while the 
electroweak preserving masses $\bar{\chi}_R \chi_L$ and $\bar{t}_R \chi_L$
may be chosen to yield the physical top mass at 175 GeV. This is the
top condensation seesaw mechanism \cite{dhseesaw}.

A consequence of this scenario is the existence of several composite
scalars. If the four-quark operators involve only the 
$\chi_R$ and the $(t,b)_L$ doublet, then the low energy theory is the 
Standard Model with a heavy Higgs boson \cite{eff}.
On the other hand, if the $t,b$ and $\chi$ participate in the 
four-quark interactions, the
composite Higgs sector includes three weak doublets and three
singlets \cite{eff}. Due to the mixing between 
the doublets, one of the neutral Higgs bosons may be light. In fact there 
is a second order phase transition from the electroweak asymmetric vacuum 
to a non-viable vacuum, in which the lightest Higgs mass vanishes.
Therefore, in this case the only lower bound on the Higgs mass is set by 
direct searches.

By contrast to other mechanisms for dynamical electroweak symmetry breaking,
this framework reduces to the Standard Model in a decoupling limit.
For example, in the three-Higgs doublet model mentioned above, one can
increase the masses of all the composite states other than the lightest 
Higgs boson, by increasing the scale of the four quark operators and the 
$\chi$ mass while keeping the VEVs close to the boundary of the second 
order phase transition. This guarantees that the models of this type are in 
agreement with the electroweak precision measurements (at least as long
as the Standard Model is in agreement).
A simpler model, with a composite Higgs sector involving only two doublets,
which reduces to the  Standard Model in the decoupling limit
is presented in Section II. 

These arguments show that the fundamental Higgs doublet from the Standard
Model may be successfully replaced by a vectorlike quark and four-quark 
interactions. Such a theory is non-renormalizable, so that one has to 
identify a suitable origin of the four-quark operators.
They may be generated by the dynamics of some spontaneously 
broken gauge group. The prototypical group of this sort is topcolor 
\cite{topcolor}, which is an embedding of $SU(3)_C$ in $SU(3) \times SU(3)$. 
Models of this type are discussed in \cite{dhseesaw,eff}. Other embeddings 
have been introduced in Ref.~\cite{family,mirror}.
A complication of this approach is that the breaking of the additional
gauge groups is non-trivial because of their strong couplings.

Another possibility is that the four-quark operators are induced by 
some gauge dynamics in compact spatial dimensions \cite{compact}. 
There are two immediate reasons that make this possibility 
attractive. First, from a four-dimensional point of view, the gauge bosons
that have a non-zero momentum in the extra dimensions appear as massive,
so that they induce four-fermion operators in the low energy theory 
without the need of breaking the gauge symmetry. Second, the gauge coupling is 
dimensionfull in more than four-dimensions such that the strength of
the gauge interactions increases with the energy, giving rise to 
non-perturbative effects \cite{tcdim}. 

In addition, the compact spatial dimensions may provide a special bonus.
To see this note that the existence of a fundamental $\chi$ quark
may be seen as artificial as the Higgs sector in the Supersymmetric Standard 
Model because its mass is not controlled by gauge invariance. However,
if the $t_R$ propagates in the extra dimensions, 
the four-dimensional fundamental $\chi$ may be replaced by 
the Kaluza-Klein excitations of $t_R$. 

The possibility of electroweak symmetry breaking due to dynamics in 
extra dimensions is discussed in section III.
%In addition, the compact spatial dimensions are required by string theory

%%%%%%%%%%%%%%%%%%%%%%%%%%%%%%%%%%%%%%%%%%%%%%%%%%%%%%%%%%%%%%%
\section{A viable composite two-Higgs-doublet model}

Certain features of the models introduced in Ref.~\cite{eff} may be 
combined to construct a minimal composite Higgs model that has the 
Standard Model as a decoupling limit. The only fundamental fields are 
the $SU(3)_C \times SU(2)_W \times U(1)_Y$ gauge bosons,
the three generations of chiral fermions, and a weak-singlet vectorlike quark, 
$\chi$, of electric charge 2/3.
Since the $\chi_L, \chi_R$ and $t_R$ transform in the same representation 
of the Standard Model gauge group, the Lagrangian includes the 
following gauge invariant mass terms 
\begin{equation}
{\cal L}_{\rm mass} = - \mu_{\chi\chi} \overline{\chi}_L\chi_R
- \mu_{\chi t} \overline{\chi}_L t_R + {\rm h.c.}
\end{equation}

An attractive interaction between $\bar{\chi}_R$ and 
$\psi^3_L \equiv (t,b)_L$ gives rise to a bound state
with the quantum numbers of a Higgs field. In order to keep the 
$\psi_L^3$ field in the low energy theory, this interaction 
has to be non-confining. 
By increasing the strength of this interaction,
the Higgs field becomes more deeply bound, until it develops a VEV.

The simplest interaction of this sort is a four-quark term in the 
Lagrangian. For breaking electroweak symmetry it is sufficient 
that the four-quark operator involves the $\chi_R$ and the $\psi_L^3$ 
\cite{eff}.
In this case, the mass of the composite Higgs boson is determined by 
the electroweak scale, and is large, close to the upper bound allowed by 
unitarity and triviality. 
On the other hand, if the $t_R$ also participates in the four-quark 
interaction, then the relation between the Higgs mass and 
the electroweak scale disappears. This is the consequence of the 
mixing between the two composite Higgs doublets that are present in 
this situation. Therefore, the Standard Model with a 
Higgs boson mass constrained from below only by direct searches, may be 
obtained in a certain decoupling limit from a theory which includes
the following four-quark operators:
\begin{equation}
{\cal L}_{\rm eff} = \frac{g_{\psi\chi}^2}{M^2} 
\left(\overline{\psi}_L^3 \chi_R \right)
\left(\overline{\chi}_R \psi_L^3  \right)
+ \frac{g_{\psi t}^2}{M^2} \left(\overline{\psi}_L^3 t_R \right)
\left(\overline{t}_R \psi_L^3 \right) ~.
\label{fourq}
\end{equation}
These interactions give rise below the scale $M$ to two composite Higgs 
doublets: $H_t \equiv \overline{t}_R \psi_L^3$ and 
$H_\chi \equiv \overline{\chi}_R \psi_L^3$. 
The Lagrangian for the composite Higgs fields\footnote{Some related 
composite two-Higgs-doublet models have been studied in \cite{twoh}.}, 
valid below the scale $M$,
includes kinetic terms involving the usual covariant derivative, 
Yukawa couplings of the scalars to their constituents, and a scalar 
potential:
\begin{equation}
{\cal L}_H = (D^\mu H_t^\dagger)(D_\mu H_t) + 
(D^\mu H_\chi^\dagger)(D_\mu H_\chi) + 
\left(\xi_t \overline{\psi}_L^3 t_R H_t
+ \xi_\chi \overline{\psi}_L^3 \chi_R H_\chi + {\rm h.c.} \right) 
+ V(H_t,H_\chi) ~.
\end{equation}
The potential, $V$, and the Yukawa couplings, $\xi_t,\xi_\chi$, can be 
determined as an expansion in $1/N_c$. Although in practice the number of 
colors is only $N_c = 3$, there is no reason to believe that the 
corrections to the leading order in $1/N_c$ change 
the results qualitatively. Furthermore, certain important features 
such as the existence of a second-order phase transition are 
independent of the $1/N_c$ expansion \cite{transition}.
In the large-$N_c$ limit, there are contributions (see Fig. 1) to the 
kinetic terms and to the following scalar terms: 
\begin{equation}
V(H_t,H_\chi) = \frac{\lambda}{2}\left[ 
(H_t^\dagger H_t)^2 + (H_\chi^\dagger H_\chi)^2 
+ 2 |H_t^\dagger H_\chi|^2 \right] + M_{H_t}^2 H_t^\dagger H_t 
+ M_{H_\chi}^2 H_\chi^\dagger H_\chi + 
\mu_{\chi\chi}\mu_{\chi t} 
\left(H_t^\dagger H_\chi + {\rm h.c.} \right) ~.
\label{potential}
\end{equation}
%%%%%%%%%%%%%%%%%%%%%%%%%%%%%%%%%%%%%%%%%%%%%%%%%%%%%%%%%%%
\begin{figure}[ht]
\begin{picture}(388,140)(-45,-20)
%%%%%%
\thicklines
\put(100,50){\circle{40}}
\thinlines
\put(80,50){\line(-2, 0){11}}
\put(63,50){\line(-2, 0){11}}
\put(46,50){\line(-2, 0){11}}
\put(95,15){$\psi^3_L$}
\put(120,50){\line(2, 0){11}}
\put(137,50){\line(2, 0){11}}
\put(154,50){\line(2, 0){11}}
\put(90,77){$t_R, \chi_R$}
%%%%
\thicklines
\put(300,50){\circle{40}}
\thinlines
\put(286,64){\line(-1,1){10}}
\put(272,78){\line(-1,1){10}}
\put(258,92){\line(-1,1){10}}
\put(290,77){$t_R, \chi_R$}
\put(290,17){$t_R, \chi_R$}
\put(265,45){$\psi^3_L$}
\put(325,45){$\psi^3_L$}
\put(286,36){\line(-1,-1){10}}
\put(272,22){\line(-1,-1){10}}
\put(258,8){\line(-1,-1){10}}
\put(314,36){\line(1,-1){10}}
\put(328,22){\line(1,-1){10}}
\put(342,8){\line(1,-1){10}}
\put(314,64){\line(1,1){10}}
\put(328,78){\line(1,1){10}}
\put(342,92){\line(1,1){10}}
\end{picture}
\caption[]{
\label{Figure}
\small  Higgs self-energy and quartic couplings induced 
in the large-$N_c$ limit.}
\end{figure}
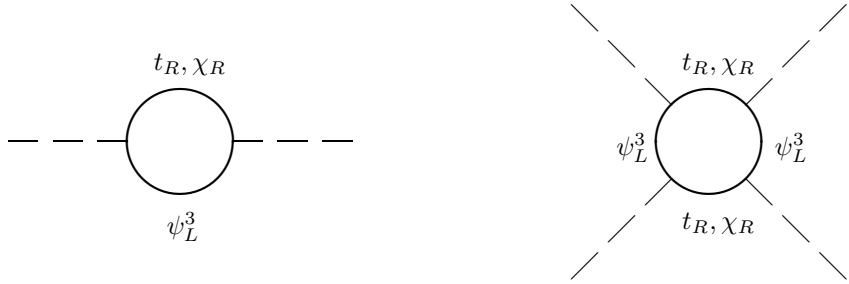
%%%%%%%%%%%%%%%%%%%%%%%%%%%%%%%%%%%%%%%%%%%%%%%%%%%%%%%%%%%
The quartic coupling $\lambda$, the scalar masses, $M_{H_t}, M_{H_\chi}$, 
and the Yukawa couplings are scale dependent parameters. 
Their values at a momentum scale $\mu$
are determined by the matching conditions at the scale $M$, namely that
all of them blow up because the Higgs fields are no longer degrees of
freedom above $M$. Also, the exchange of the Higgs fields
should give rise to the four quark interactions (\ref{fourq}), so that 
at the scale $M$
\begin{equation}
\frac{g_{\psi t}^2}{M^2} = \frac{\xi_t^2}{M_{H_t}^2} \;\;\; , \;\; \;\;
\frac{g_{\psi \chi}^2}{M^2} = \frac{\xi_\chi^2}{M_{H_\chi}^2} \; ~.
\end{equation}
Computing the diagrams of Fig.~1 with a momentum cut-off $M$,
gives for $\mu \ll M$ the following results \cite{eff}: 
\begin{equation}
\xi_t = \xi_\chi = \sqrt{\frac{\lambda}{2}} =
\frac{4 \pi }{ \sqrt{N_c \ln \left(M^2/\mu^2\right)}} \; ~,
\label{yuk}
\end{equation}
\begin{equation}
M_{H_{t,\chi}}^2 = \frac{2 M^2}{\ln \left(M^2/\mu^2\right)}
\left(\frac{8\pi^2}{N_c g^2_{\psi t, \chi}} - 1 \right) 
+ 2 \mu_{\chi\chi}^2 ~.
%\frac{\mu_{\chi\chi}^2}{M^2} \ln \left( \frac{M^2}{\mu^2}\right) \right] ~.
\label{scmass}
\end{equation}
From the expression for the masses it follows that as the four-quark
couplings $g^2_{\psi t, \chi}$ are varied, the vacuum suffers a 
second order phase transition (or at least a weakly first order one).
Note that in addition to the terms displayed in Eq.~(\ref{potential}), 
the potential includes higher dimensional terms and 
sub-leading terms in $\mu/M$, which may be neglected.
The $H_t^\dagger H_\chi$ term is appropriately included in $V$ because
for $g^2_{\psi t, \chi}$ close to the critical value, the 
$|M_{H_{t,\chi}}^2|$ squared masses may be comparable or smaller than 
$\mu_{\chi\chi}\mu_{\chi t}$.
In the presence of this term, the two doublets acquire aligned 
VEVs for a range of  $g^2_{\psi t, \chi}$:
\begin{equation}
\langle H_t \rangle = \frac{1}{\sqrt{2}} 
\left( \begin{array}{c} v \cos\beta \\ [2mm] 0 \end{array} \right)
\;\;\; \;\; , \;\; \;\;\;\; 
\langle H_\chi \rangle = \frac{1}{\sqrt{2}} 
\left( \begin{array}{c} v \sin\beta \\ [2mm] 0 \end{array} \right) ~.
\end{equation}
where the electroweak scale, $v\approx 246 \; {\rm GeV}$,
and $\tan \beta$ are determined from the minimization conditions 
as functions of $M_{H_{t,\chi}}^2/\lambda$ and 
$\mu_{\chi\chi}\mu_{\chi t}/\lambda$.
The top quark mass is given by the smaller eigenvalue of the 
$t - \chi$ mass matrix:
\begin{equation}
\left( \begin{array}{cc} \frac{1}{\sqrt{2}} \xi_t v \cos\beta \; & 
\;  \frac{1}{\sqrt{2}} \xi_\chi v \sin\beta \\ [2mm] 
		\mu_{\chi t} \; & \; \mu_{\chi\chi} \end{array} \right)
\end{equation}
For example, when $\tan \beta \gg 1$ the top mass is given by a seesaw 
relation:
\begin{equation}
m_t \approx \frac{\xi_\chi v}{\sqrt{2}} \frac{\mu_{\chi t}}{\mu_{\chi\chi}} ~.
\end{equation}
Note that $\xi_\chi v/\sqrt{2} \sim 0.5$ TeV, as mentioned in Section I.
The lightest CP-even neutral Higgs boson has a squared-mass
\begin{equation}
M_{h^0}^2 = \lambda v^2 + M_{H_{t}}^2 + M_{H_{\chi}}^2 
- \left[
\left(\frac{\lambda v^2}{2} \cos 2\beta + 
M_{H_{t}}^2 - M_{H_{\chi}}^2 \right)^2 + 
\left(\frac{\lambda v^2}{2} \sin 2\beta + 
2 \mu_{\chi t}\mu_{\chi\chi}\right)^2 \right]^{1/2} \; ~.
\end{equation}

Consider the following decoupling limit 
\begin{equation}
M_{H_{t}}, M_{H_{\chi}}, \mu_{\chi\chi} \gg \mu \sim v ~.
\end{equation}
This limit implies a fine-tuning of the mass parameters in the 
potential such that the two VEVs remain fixed when the masses
increase. 
Eqs.~(\ref{yuk}) and (\ref{scmass}) 
remain valid provided $M \gg \mu_{\chi\chi}^2/\mu$.
It is clear that a tuning of $M_{H_{t,\chi}}$, $\mu_{\chi\chi}$ and 
$\mu_{\chi t}$ allows the $h^0$ to remain light, while the other 
%\begin{equation}
%M_{H_{t}}^2 M_{H_{\chi}}^2 - \mu_{\chi\chi}^2\mu_{\chi t}^2 
%\end{equation}
physical scalar masses decouple.
Therefore, in this decoupling limit the composite Higgs sector
reduces to the Standard Model Higgs.
In actuality it is not necessary to tune the parameters
more finely than a few percent in order to keep the corrections
to the electroweak observables in agreement with the data
(the main constraint comes from the $\rho$ parameter 
\cite{dhseesaw,eff,isospin}).

With the electroweak symmetry broken correctly and the top quark
mass accommodated as shown above, it remains to produce the masses 
and mixings of the other quarks and leptons. In the effective theory 
below the scale $M$, these are easily generated by perturbative
four-fermion operators which induce Standard Model Yukawa couplings.
For example, the operators 
\begin{equation}
\frac{1}{M^2} \left( \overline{\chi}_R \psi_L^3\right) 
\left( \overline{l}_L^j i\sigma_2 e_R^k\right) 
\label{yukmas}
\end{equation}
induce Yukawa couplings between $H_\chi$ and the charged 
lepton fields, $l_L^j$ and $e_R^i$, ($j,k$ are generational indices). 
These four-fermion operators and the non-perturbative four-quark operators 
that produce the Higgs bound states have to originate in
physics above the scale $M$. 
Another possibility for fermion mass generation is to extend the 
seesaw mechanism to all the quarks and leptons \cite{family}.

%%%%%%%%%%%%%%%%%%%%%%%%%%%%%%%%%%%%%%%%%%%%%%%%%%%%%%%%%%%%%%%
\section{Gauge Dynamics in Compact Spatial Dimensions}

The non-perturbative four-quark operators introduced in Section II 
may arise naturally in the presence of compact spatial dimensions 
\cite{compact}.
Although this is a rather general statement, it is instructive to discuss 
it from the perspective of string theory. 

String (or M) theory predicts the existence of 16 or 
32 conserved supercharges, and six
(seven) extra spatial dimensions at the string scale $M_s$. 
It has been traditionally assumed that i) $M_s$ is close to the 
Planck scale, ii) the extra dimensions are compactified at $M_s$, 
and iii) $N=1$ supersymmetry is preserved all the way 
down to the electroweak scale.
All these assumptions are consistent with a perturbative string 
picture. However, the recent progress in understanding 
non-perturbative string dynamics raises questions about these three
assumptions. 

First, it is natural that the string scale $M_s$ is lowered down to the 
GUT scale \cite{witten}, and in fact it can be placed anywhere below 
the Planck scale, with a lower bound set by phenomenology in the TeV 
range \cite{lykken}. A remarkable realization of this idea \cite{largedim}
is based on the observations that there may be compact dimensions as large as 
1 mm provided they are accessible only to gravitons, and that 
the scale where gravity becomes strong 
is suppressed by the volume of the compact space.
The restriction of matter and gauge fields to a 
surface in extra dimensions is permitted in string and M theory
by the existence of branes.

Second, extra dimensions accessible to both gravitons and gauge bosons
may also have compactification scales as small as a few TeV \cite{kkstates},
and in principle could be somewhat larger than $M_s^{-1}$.

Finally, non-perturbative string effects may break supersymmetry completely 
at the $M_s$ scale. For example, a general manifold
does not preserve any supercharge, so that if some of the spatial dimensions
have the compactification scale at $M_s$, the low energy theory may be 
a non-supersymmetric field theory.

Of course, string theory is still plagued with phenomenological
disasters, such as the cosmological constant problem, the existence of 
potentially massless moduli, or the inability of predicting the 
Standard Model at low energy. However, these problems might be solved in the 
future. Until the non-perturbative string effects
will be better understood, it is useful to adopt a phenomenological approach,
and to investigate various scenarios for physics beyond the Standard Model 
inspired by the above considerations regarding string theory.

A commonly used assumption for allowing chiral fermions 
in the four-dimensional theory obtained upon dimensional reduction,
is that a quantum field theory may be defined 
on an orbifold, which is a space with singularities.
This assumption, made a long time ago \cite{kkstates} and 
reinvigorated by the $S^1/Z_2$ compactification of M theory
\cite{hora}, has been used to study supersymmetry breaking
\cite{peskin}, gauge coupling 
unification \cite{ddg}, and other phenomenological and theoretical
issues \cite{flavor,extrad} regarding extra dimensions.  
It is not clear whether the string dynamics may be treated below  $M_s$
as pure quantum field theory in a higher dimensional space, especially 
since the separation between the compactification scale and $M_s$ is not 
large, but one can assume that this is the case and investigate 
the consequences.

The above considerations have a striking connection
with the composite Higgs models: the required four-quark interactions may be 
induced by gauge dynamics in extra dimensions.
The physical picture employed here is that the eleven-dimensional spacetime
of M theory includes three flat spatial dimensions and seven compact
dimensions with a sufficiently large volume to 
allow $M_s$ much below the Planck scale.
The gauge fields are restricted to a region of this space which includes
the three-dimensional flat space and has a thickness larger than 
$M_s^{-1} $ in $\delta$ extra dimensions. 
In this case, the gluons that have zero momentum in the extra dimensions
correspond to the massless QCD gluons, while the ones with non-zero 
momentum in the extra dimensions are massive gauge bosons from a 
four-dimensional point of view, with a spectrum given by  
\begin{equation}
M_{n_1, ..., n_\delta}^2 = \sum_{l = 1}^{\delta}\frac{n_l^2}{R_l^2} ~,
\end{equation}
where $R_l$ are the compactification radii of the $\delta$ extra 
dimensions, and $n_l$ are the Kaluza-Klein (KK) excitation numbers.
If the quarks propagate only on the 
three-dimensional boundary of the $3 + \delta$ space accessible to 
the gluons (this corresponds to the fixed point of an orbifold),
then the couplings of the KK modes of the gauge bosons 
to the quarks and leptons are identical (up to an overall normalization) 
with those of the Standard Model gauge bosons.
Therefore, the tree level exchange of the gluonic KK modes induces 
flavor universal four-quark operators
in the low energy theory: 
\begin{equation}
{\cal L}_{\rm eff}^c =
- \sum_{l=1}^{\delta} \sum_{n_l \ge 0}
\frac{g_s^2}{2 M_{n_1, ..., n_\delta}^2} 
\left(\sum_{q} \overline{q} \gamma_\mu T^a q \right)^2 ~,
\label{ops1}
\end{equation}
where $q$ are all the quark fields, $T^a$ are the $SU(3)_{\rm C}$
generators, and $g_s$ is the QCD gauge coupling.
The sum over KK modes should be cut off at modes of mass $\sim M_s$.
The left-right current-current part of ${\cal L}_{\rm eff}^c$ 
has the same form (due to a Firz transformation and the large-$N_c$ limit)
as the operators that produce scalar bound states,
while the other parts of ${\cal L}_{\rm eff}^c$ do not contribute 
in the large-$N_c$ limit to the scalar potential.

It is clear that for a sufficiently large number of extra dimensions, 
the number of KK modes, $N_{KK}$, 
grows to the point where the attractive interaction 
between the quarks is strong enough to trigger electroweak symmetry breaking.
In practice, the sum (\ref{ops1}) is super-critical for $\delta = 4$ 
dimensions of size $\sim 2/M_s $. 
This result \cite{compact} is derived at tree level, so that
it may be modified when loops are included. However, the qualitative picture
appears to be robust.

Since the gluon couplings are flavor universal, all the quarks participate 
as constituents in scalar bound states. The VEVs of these scalars may be 
controlled by additional interactions. For example, the KK modes of the 
hypercharge gauge boson give rise to four-quark interactions which are 
attractive for the up-type quarks and repulsive for the down-type ones.
Therefore, the scalars made up of up-type quarks are the most strongly 
bound and they will be the only ones with non-zero VEVs provided the 
flavor universal interactions are close to criticality.
Furthermore, some flavor non-universal interactions should prevent
the bound states involving the up and charm quarks from developing 
large VEVs. Such interactions, as well as the perturbative 
four-fermion terms that should induce the light quark and lepton masses,
could be accommodated by various flavor physics scenarios above 
the compactification scale \cite{flavor}.

It is convenient to assume that $M_s$ corresponds to the scale where 
the perturbative expansion in $\alpha_s N_{KK}$ breaks down,
which is usually larger than the compactification scale 
by less than one order of magnitude.
Given that the compactification scale corresponds to the scale $M$ 
of the four-quark operators, which is most likely in the 
$10 - 50$ TeV range, the string scale is expected to be 
$M_s \sim {\cal O}(100)$ TeV. The compact space accessible to the 
gravitons has in this case a seven-dimensional volume 
of order (10 GeV)$^{-7}$.

The framework discussed here can be extended in different directions.
For example, instead of using the gluonic KK modes for binding the 
composite scalar states, one can use some new gauge symmetries.
Most economical would be an anomalous $U(1)$ which leads to the 
formation of only two Higgs doublets, as in Section II.
Another interesting possibility is to allow the $t_R$ to propagate 
in extra dimensions such that its KK modes play the role of $\chi$.
Some of the above arguments change in this case because the 
couplings of the gluonic KK modes are modified.

In conclusion, a composite Higgs sector may be provided by quark-antiquark
states bound by gluons propagating in compact dimensions.
If the Higgs sector will be discovered in collider experiments, then 
the study of the spectrum will allow to distinguish the origin of 
the attractive four-quark interactions: new gauge bosons, or extra dimensions.

%%%%%%%%%%%%%%%%%%%%%%%%%%%%%%%%%%%%%%%%%%%%%%%%%%
%\section{Conclusions}
%\begin{eqnarray}
\vspace*{3mm}

{\it Acknowledgements:} I would like to thank Chris Hill and Nima Arkani-Hamed
for illuminating conversations, and the organizers of DPF'99 for a stimulating
environment. 
Fermilab is operated by the URA under DOE contract DE-AC02-76CH03000.

%%%%%%%%%%%%%%%%%%%%%%%%%%%%%%%%%%%%%%%%%
 
\end{document}